\documentclass[iop]{emulateapj}

\usepackage{epsfig}
\usepackage{natbib}
\def\simgt{\lower.5ex\hbox{$\; \buildrel > \over \sim \;$}}
\def\simlt{\lower.5ex\hbox{$\; \buildrel < \over \sim \;$}}

\def\amin{\ifmmode^{\prime}\else$^{\prime}$\fi}
\def\asec{\ifmmode^{\prime\prime}\else$^{\prime\prime}$\fi}

\def\simgt{\lower.5ex\hbox{$\; \buildrel > \over \sim \;$}}
\def\simlt{\lower.5ex\hbox{$\; \buildrel < \over \sim \;$}}

\newcommand\chandra{{\it Chandra}}

\newcommand\nustar{{\it NuSTAR\/}}

\def\sgra{Sgr A*}

\slugcomment{Accepted for publication to The Astrophysical Journal Letters}

\shorttitle{NuSTAR Detection of Cannonball}
\shortauthors{}

\begin{document}

\title{High-Energy X-rays from J174545.5-285829, the Cannonball: \\ A Candidate Pulsar Wind Nebula Associated with Sgr~A~East}

\author{
Melania Nynka\altaffilmark{1}, Charles J. Hailey\altaffilmark{1}, Kaya Mori\altaffilmark{1}, Frederick K. Baganoff\altaffilmark{2}, 
Franz E. Bauer\altaffilmark{3,4}, Steven E. Boggs\altaffilmark{5}, William W. Craig\altaffilmark{5,6}, 
Finn E. Christensen\altaffilmark{7}, Eric V. Gotthelf\altaffilmark{1}, Fiona A. Harrison\altaffilmark{8}, 
Jaesub Hong\altaffilmark{9}, Kerstin M. Perez\altaffilmark{10}, Daniel Stern\altaffilmark{11}, 
Shuo Zhang\altaffilmark{1}, and William W. Zhang\altaffilmark{12}} 

\altaffiltext{1}{Columbia Astrophysics Laboratory, Columbia University, New York, NY 10027, USA}
\altaffiltext{2}{Kavli Institute for Astrophysics and Space Research, Massachusetts Institute of Technology, Cambridge, MA 02139, USA}
\altaffiltext{3}{Instituto de Astrof\'{\i}sica, Facultad de F\'{i}sica, Pontificia Universidad Cat\'{o}lica de Chile, 306, Santiago 22, Chile}
\altaffiltext{4}{Space Science Institute, 4750 Walnut Street, Suite 205, Boulder, CO 80301, USA}
\altaffiltext{5}{Space Sciences Laboratory, University of California, Berkeley, CA 94720, USA}
\altaffiltext{6}{Lawrence Livermore National Laboratory, Livermore, CA 94550, USA}
\altaffiltext{7}{DTU Space - National Space Institute, Technical University of Denmark, Elektrovej 327, DK-2800 Lyngby, Denmark}
\altaffiltext{8}{Cahill Center for Astronomy and Astrophysics, California Institute of Technology, Pasadena, CA 91125, USA}
\altaffiltext{9}{Harvard-Smithsonian Center for Astrophysics, Cambridge, MA 02138, USA}
\altaffiltext{10}{Columbia University, New York, NY 10027, USA}
\altaffiltext{11}{Jet Propulsion Laboratory, California Institute of Technology, Pasadena, CA 91109, USA}
\altaffiltext{12}{NASA Goddard Space Flight Center, Greenbelt, MD 20771, USA}

\begin{abstract}

  We report the unambiguous detection of non-thermal X-ray emission up to
  $30$~keV from the Cannonball, a few-arcsecond long diffuse X-ray feature near the Galactic Center, using the \nustar\ X-ray observatory.
  The Cannonball is a high-velocity ($v_{proj} \sim 500$~km~s$^{-1}$)
  pulsar candidate with a cometary pulsar wind nebula (PWN)
  located $\sim 2'$ north-east from \sgra, just outside the radio shell of the supernova remnant Sagittarius~A (Sgr~A) East. Its non-thermal X-ray spectrum, measured
  up to $30$~keV, is well characterized by a $\Gamma\sim1.6$ power-law, typical of a PWN, and has an X-ray
  luminosity of $L(3-30\ {\rm keV}) = 1.3 \times
  10^{34}$~erg~s$^{-1}$. The spectral and spatial results derived from
  X-ray and radio data strongly suggest a runaway neutron star
  born in the Sgr~A~East supernova event. We do not find any pulsed signal from the Cannonball.  The \nustar\ observations
  allow us to deduce the PWN  magnetic field and show that it is consistent
  with the lower limit obtained from radio observations.

\end{abstract}

\section{Introduction}

Sagittarius A (Sgr~A) East is an elongated non-thermal radio shell, which in the west encompasses \sgra\ and Sgr~A West, and in the east is bounded 
by a molecular cloud with which it is likely interacting. 
The region interior to the radio shell emits thermal X-rays.
\citet{Park2005} (hereafter P05), using data from \chandra\ observations, found the X-ray emission comes from three distinct regions:
a Center region highly enriched in Fe and other elements; a North region more modestly enriched in Fe;  and a `Plume' region with solar abundances.
Each region contains a thermally emitting plasma of temperature $\sim1$~keV. 
The North region contains a higher temperature component with $kT\sim5-10$~keV.
Based on the total Fe abundance, several authors conclude that Sgr~A East is likely the result of a Type~II supernova~(SN) explosion 
\citep{Maeda2002, Sakano2004, Park2005}.
However, none of the observations conclusively rule out a Type Ia explosion. 
The presence of a non-thermal radio shell with no corresponding X-ray emission, along with the center-filled X-ray morphology, argue persuasively that 
Sgr A~East is a mixed morphology supernova remnant (SNR) \citep{Maeda2002}. 

Outside the radio shell of Sgr A~East, just north of the Plume, lies
the \chandra\ source CXOGC J174545.5$-$285829 \citep{Muno2003}, termed `the Cannonball.'  This
source has faint surrounding asymmetric X-ray
emission, slightly elongated in the north-south direction, suggesting a
cometary tail pointing back to the approximate center of Sgr~A East
(P05).  Based on its non-thermal X-ray emission, spatial extent ($\sim
0.1$~pc at $8$~kpc), association with Sgr~A East, and lack of flux
variation on year time scales, CXOGC~J174545.5$-$285829 has been
proposed as a runaway neutron star (NS) with a pulsar wind nebula (PWN).  
However, due to the lack of spectral coverage above 8~keV, \chandra\ was unable to resolve a high-temperature thermal model, 
such as what would be found in a cataclysmic variable, from a non-thermal spectrum typical of a PWN.
So far, pulsations remain undetected.

The radio observations present a similar picture. A radio counterpart
to the Cannonball is detected at 5.5~GHz \citep*[hereafter ZMG13]{zmg2013}  
that connects to a radio plume region ($\sim30\asec~\times~15\asec$) and a long radio tail
($\sim30\asec$) trailing from the radio shell back into the interior
of Sgr~A~East. This morphology is similar to that observed for several
runaway NSs \citep[e.g. the Mouse;][]{Gaensler2004}. 
The proper motion, implying a transverse speed of
$\sim500 \pm 100$~km~s$^{-1}$, flat spectrum ($\alpha=-0.44 \pm 0.08$ for
the compact head, $\rm{S}_{\nu}\propto\nu^{- \alpha}$ ) and the cooling linear tail ($\alpha=-1.94\pm0.02$; ZMG13) 
all strongly suggest that the Cannonball is a PWN associated with a NS 
that has overrun the shell of Sgr~A East.

Here we report the \nustar\ discovery of non-thermal X-rays from the
Cannonball up to energies $30$~keV.  In $\S2$  we describe the
\nustar\ observations. In $\S3$, $\S4$ and $\S5$ we present the imaging, spectral 
and timing analysis, respectively.  In $\S6$,
we use \nustar\ and \chandra\ data to estimate the PWN magnetic
field strength independent of the lower limit derived from the radio observations of ZMG13.

\section{\nustar\ observations}

\nustar\ is the first focusing telescope that operates in the hard
X-ray band of $10-79$~keV.  It consists of two coaligned optics/detector pairs, focal plane modules A~and~B (FPMA and FPMB),
and has a field of view of $10\amin \times 10\amin$ at $10$~keV.
\nustar\ has an angular resolution of 18\asec\ full width at half maximum (FWHM) and 58\asec\ half
power diameter, and an energy resolution of 400~eV (FWHM) at 10~keV
\citep{Harrison2013}. The nominal \nustar\ timing resolution is 4~ms.
The nominal \nustar\ reconstruction
coordinates are accurate to $8''$ ($90\%$ confidence level). In the
following study, all observation were processed using {\it nupipeline}, 
NuSTARDAS v. 1.1.1, which is used to generate
response matrices and exposure maps, and the HEASoft v. 6.13 package
is used for data analysis. 

\nustar\ observed Sgr~A* three times between 2012~July~20 and 2012~October~10 \citep{Barrier2013}. In addition, 
\nustar\ triggered multiple target of opportunity observations(ToO), four of which were used in the subsequent analysis, of the
outburst from the newly-discovered magnetar SGR~J1745-29 near Sgr~A*
in 2013 \citep{Mori2013}. The Cannonball, located $\sim2\amin$ away
from Sgr~A*, was captured in the \nustar\ field of view in all 
seven observations, with a total exposure time of 429~ks (Table \ref{tab:obslog}). 
We performed imaging and spectral analysis using
the three Sgr A* observations, as the Cannonball falls on or near
the detector chip gaps in the SGR~J1745$-$29 observations, which were
otherwise suitable for timing analysis.  Photon arrival
times were corrected for on-board clock drift and precessed to the
solar system barycenter using the JPL-DE200 ephemeris and the
\chandra\ subarcsec position (P05).

\begin{deluxetable}{lcccc}
\tablecaption{\nustar\ Log of Cannonball Observations}
\tablewidth{0pt}
\tablecolumns{4}
\tablehead { \colhead{ObsID}    &   \colhead{Start Date}   &   \colhead{Exposure}   &  \colhead{Target} \\
 \colhead{     }  & \colhead{(UTC)}        & \colhead{(ks)}     }
\startdata
30001002001  & 2012 07 20 & 166.2    & Sgr~A*\\
30001002003  & 2012 08 04 &   83.8   & Sgr~A*\\
30001002004  & 2012 10 16 &   53.6   & Sgr~A*\\
80002013002  & 2013 04 27 &   54.1   & Magnetar ToO\\
80002013004  & 2013 05 04 &   42.1   & Magnetar ToO\\
80002013006  & 2013 05 11 &   35.6   & Magnetar ToO\\
80002013012  & 2013 06 14 &  29.1    & Magnetar ToO\\
\enddata
\tablecomments {The exposure times listed are corrected for good time intervals.}
\label{tab:obslog}
\end{deluxetable}     

\section{Imaging analysis}

\begin{figure*}[t]\centering
       \psfig{figure=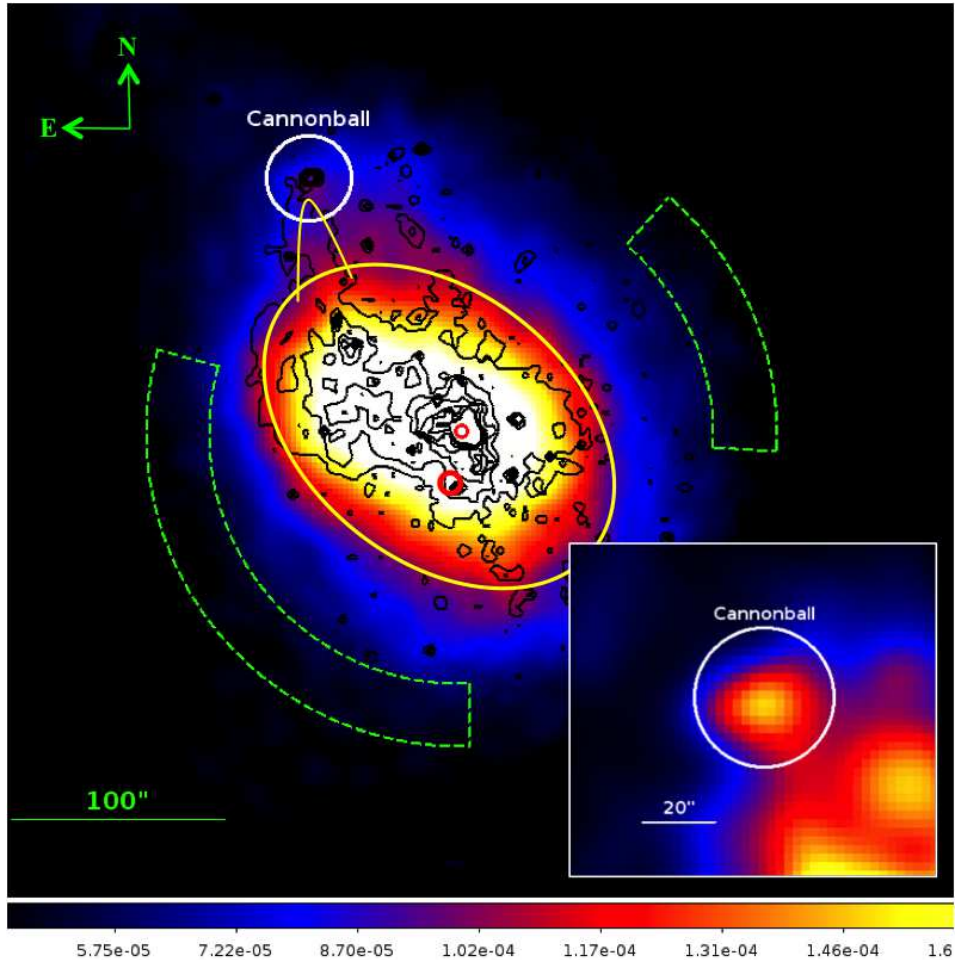, width=0.7\linewidth, angle=0}
\caption{
\nustar\ mosaic image of the Cannonball, Plume, and SgrA* region in $3-79$~keV band. The Cannonball is indicated by the white circle with radius of $20''$. 
Background regions are indicated by the green dashed regions. \chandra\ $3-8$~keV contours are overlaid in black, and approximate positions of the Plume and 
SgrA East are in yellow. The image is shown with a linear color scale 
and the scale range chosen to highlight the Cannonball emission.  Inset-- \nustar\ mosaic image of the Cannonball in $10-30$~keV band.
}
\label{fig:img}
\end{figure*}

We applied astrometric corrections to each \nustar\ event file by aligning \nustar-detected objects 
with their reference \chandra\ locations \citep{Muno2003}.
This is particularly important in the crowded Galactic Center region as it increases the significance
of faint sources.
We then generated a mosaiced \nustar\ image of the Cannonball by merging
exposure-corrected images from each observation.  The resulting
$3-79$~keV mosaic is shown in Figure \ref{fig:img}, along with the
$3-8$~keV \chandra\ contours (P05).  It is clear that the \chandra\ contours are
centered on the bright Cannonball emission in the \nustar\
image.  The $10-30$~keV \nustar\ image is shown as an inset.
The Cannonball is detected as an isolated source at
energies above 10~keV.
Above 30~keV the source becomes background-dominated, and below 10~keV the source is too faint to be detected over thermal
emission from the Plume region and Sgr~A~East, which contribute due to the \nustar\ PSF.

We investigated whether the Cannonball exhibits any elongation above 10~keV as observed in the radio band (ZMG13). First, we generated an effective PSF by taking into 
account the appropriate off-axis angle and orientation for each observation then adding individual PSFs weighed by exposure time. Second, we used the effective 
PSF to convolve a 2-D Gaussian model profile to fit a circular region with $30\asec$ radius 
around the Cannonball using {\tt Sherpa}, a CIAO modeling and fitting application \citep{Fruscione2006}. 
The best-fit centroid position of the Cannonball  overlaps the \chandra\ position within the position uncertainty of $3.3\asec$ (90\% confidence). 
The best-fit FWHM=3.0$''\pm2.7''$ shows that the Cannonball is not resolved by \nustar.
Additionally, we analyzed residuals of the 2-D image by fitting along various axes through the Cannonball, and did not find any elongation 
of the source within statistical uncertainties.  
We cannot detect the small $\sim2''$ tail observed by \chandra\ (P05).

\section{Spectral analysis}

We extracted source spectra from the three \nustar\ Sgr~A* observations using a $20\asec$ radius circular region centered on the \chandra\ position 
(white circle in Figure \ref{fig:img}). This region is optimal for spectral analysis since it significantly reduces the contribution of the nearby 
Plume and Sgr~A~East emission, indicated in Figure \ref{fig:img}, and maximizes the signal-to-noise ratio. 
After rebinning \nustar\ spectra to $>20$ counts per bin, we performed spectral fitting with {\tt XSPEC} version 12.8.0 \citep{Arnaud1996} and adopted the abundance and 
atomic cross-section data from \citet{Wilms2000} and \citet{Verner96}, respectively.
Due to the broad \nustar\ PSF, this source extraction region will also contain thermal emission from the Sgr~A~East Plume and North regions. 
The \nustar\ spectrum can therefore be fit with an absorbed thermal and non-thermal model, {\bfseries Tbabs*(apec+powerlaw)}.

\nustar\ background spectra were extracted from partial annuli shown in Figure \ref{fig:img} (green dashed region).  
These regions were chosen to remove instrument and cosmic X-ray background, as well as the contribution of point sources 
detected by \chandra\  \citep{Muno2003}.  These sources,
which are too faint to be detected by \nustar\ individually, collectively add an additional background component.  
\citet{Muno2003} used \chandra\ to resolve these point sources and determined that they were peaked at
the Galactic Center and decrease radially from Sgr~A* outward.  We therefore chose 
background regions at the same radial distance from Sgr~A* as the Cannonball, and avoided the known filaments, Sgr~A~East and detector chip gaps. 

We used data from \chandra\ ACIS-I to constrain the column density. 
The \chandra\ spectrum was obtained from 82 ACIS-I observations, spanning from 2000~October to 2012~October, with a total integration time of 4~Ms.
We used a source region $r=1.6''$ to remove contamination from the adjacent soft foreground star, CXOGC~J174545.2-285828, which has a thermal spectrum 
negligible above 2~keV (P05) and does not affect \nustar\ spectra.  
The  \chandra\ background spectra were extracted from a region adjacent to the Cannonball, 
in order to subtract thermal emission from the Plume and Sgr~A~East.
The \chandra\ spectrum can be fit with an absorbed power-law model, {\bfseries Tbabs*powerlaw}.

\begin{figure*}[t]
 \centerline{ \hfill
       \psfig{figure=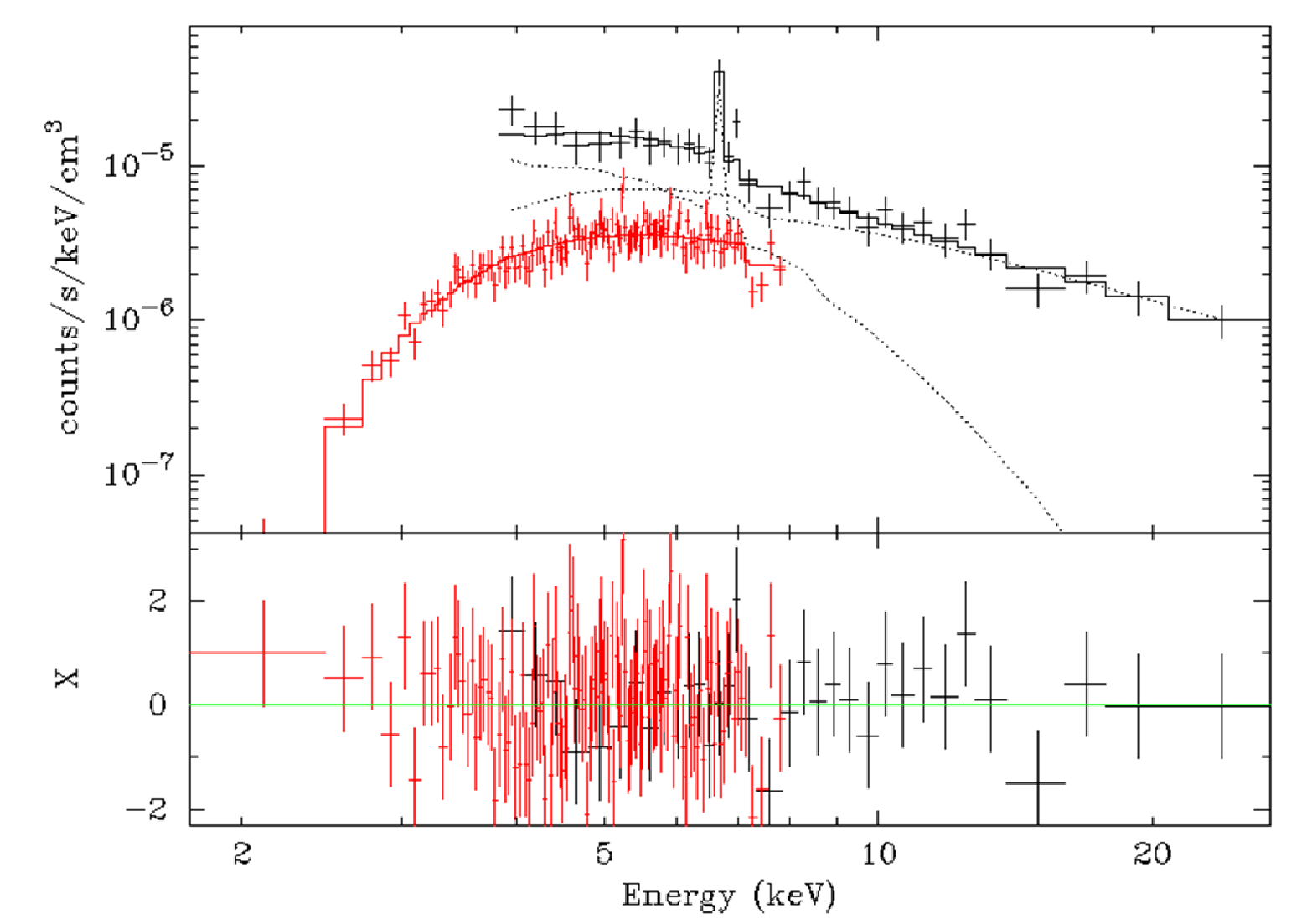,width=0.9\linewidth, width=0.5\linewidth, angle=0} \hfill
       \psfig{figure=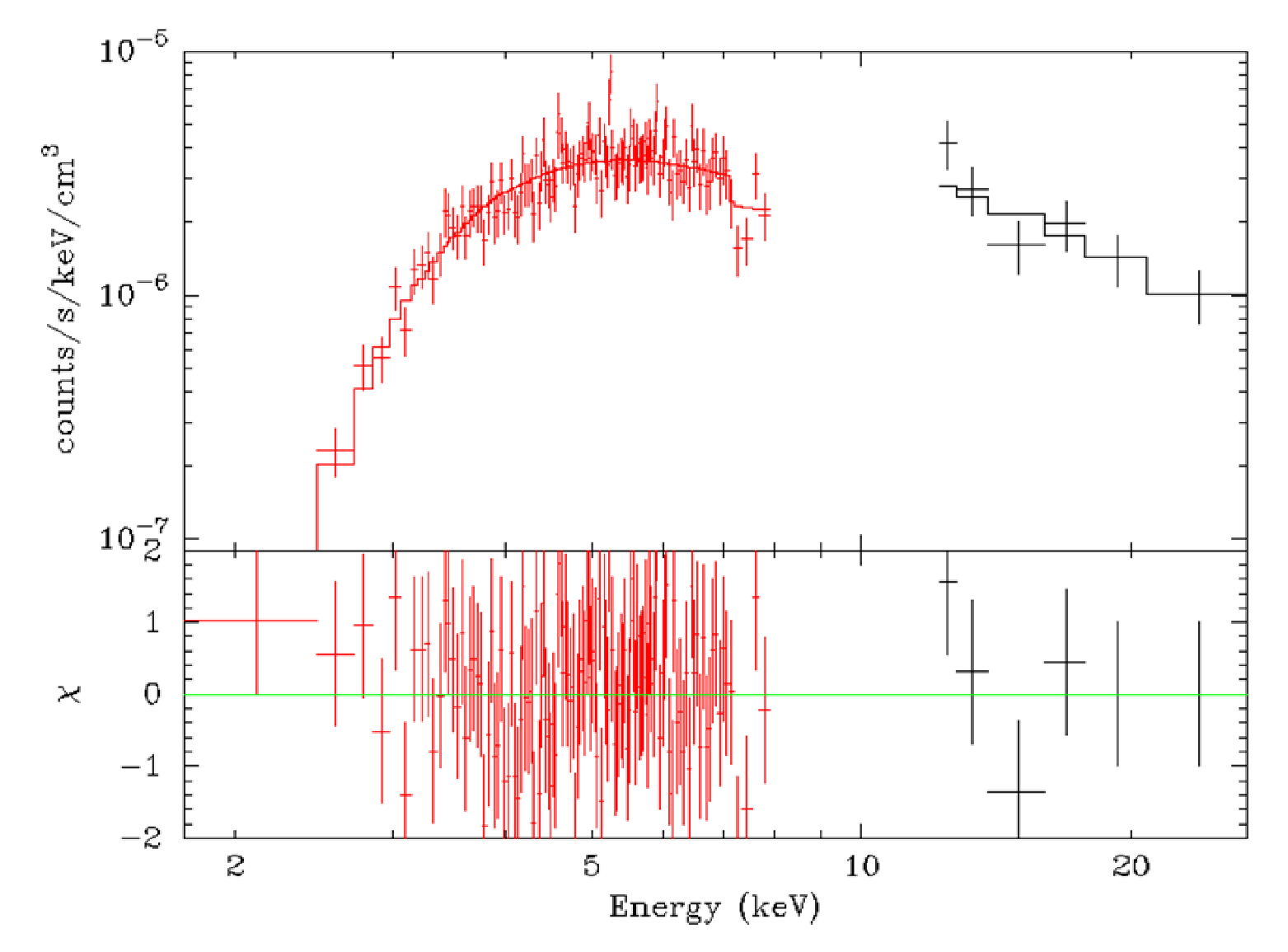,width=0.9\linewidth, width=0.5\linewidth, angle=0} \hfill }
\caption{
Cannonball joint fit with \nustar\ and \chandra. {Left--} 2-30~keV joint fit.  Solid lines are the best-fit
model.  Dashed lines are the model components (thermal, and non-thermal) of the \nustar\ data. {Right--}  Cannonball-only spectral fit with an absorbed power-law model.
For both images, black indicates \nustar\ FPMA data, red indicates \chandra\ {\it ACIS} data. 
}
\label{fig:spectra}
\end{figure*}

We then jointly fit the \nustar\ and \chandra\ data.  
The absorption coefficient and the power-law
parameters were linked between the two data sets.  We analyzed the \nustar\ spectra in  $3-30$~keV and  
\chandra\ data in $2-8$~keV.
The fit results are shown in Figure \ref{fig:spectra} and Table \ref{tab:spectable}.
The best-fit power-law photon index is
$\Gamma=1.6\pm0.4$, consistent with the \chandra\ results of P05.
The discrepancy between the best-fit absorption column and that found in P05 is due to our use of more recent 
absorption and abundance data.

As an alternative approach, we restricted the \nustar\ data to the $13-30$~keV band to cleanly isolate the non-thermal Cannonball emission from the low-temperature 
thermal components of the Sgr~A~East Plume and North regions.
We then jointly fit 
\nustar\ and \chandra\ spectra with an absorbed power-law model for both spectra, with all parameters linked. The best-fit parameters for the power-law model, shown in Figure 
\ref{fig:spectra} and Table \ref{tab:spectable}, are consistent with the values found in the full-band spectral analysis.

\begin{deluxetable}{lcccccc}
\tablecaption{Results from NuSTAR + Chandra joint spectral analysis}
\tablewidth{0pt}
\tablecolumns{3}
\tablehead { \colhead{Parameters}  & \colhead{Power-law+Apec} & \colhead{Power-law} }
\startdata
\nustar\ Energy Band [keV]          & $3-30$                 & $13-30$ \\
$N_{\rm H} [10^{22}$\,cm$^{-2}$]    & $31^{+2}_{-3}$         & $32^{+2}_{-3}$  \\
Power-law Index                     & $1.6^{+0.3}_{-0.4}$    & $1.6^{+0.4}_{-0.3}$ \\
kT  [keV]                           & $2.8\pm0.1$            & $\cdots$ \\
$\chi_\nu^2$ [dof]                  & 0.81 (66)              & 0.90 (39) \\
$F_{\rm X}(2-10$~keV)               & $7\pm3$                & $7\pm2$ \\
$F_{\rm X}(10-30$~keV)              & $9^{+2}_{-3}$          & $9\pm1$ \\                                                            
\enddata                                                                                                                                                                                               
\tablecomments { Best-fit parameters of the Cannonball, joint fit  \nustar\ and \chandra\ data.  The errors are $90\%$ confidence level. 
Flux values are obtained from the power-law component, and have units $10^{-13}$\,erg\,s$^{-1}$\,cm$^{-2}$.  }                                                     
\label{tab:spectable}                                                                                                                  
\end{deluxetable}


\section{Timing Analysis}

The high time resolution of \nustar\  allows a search for pulsations 
down to periods of $P=4$~ms, covering the expected range for a rotation powered pulsar.
Previous searches were restricted to $P>147$~ms (P05).
We evaluated the power
at each frequency (oversampling by a factor of 2) using the $Z^2_n$ test statistic for $n=1,2,3,5$, to
be sensitive to both broad and narrow pulse profiles. We initially restricted the timing search to photons
energies in the $3-25$~keV range and used an aperture of
$20^{\prime\prime}$ to optimize the signal-to-noise ratio.

From a search of all the observations, the most significant signal was
$Z^2_3 = 48.61$ for ObsID 30001002001, corresponding to a probability of
false detection of $\wp = 0.25$ for
$28\times(8.9\times10^9)$ search trials. The resulting period is not reproduced in the other observations.  We
repeated our search for an additional combination of energy ranges
$3<E<10$~keV, $10<E<25$~keV and aperture size $r<10^{\prime\prime}$.
We also searched the combined data set in ($f,\dot f$)-space around our best candidates from the individual observations.
None of these resulted in a significant detection.
We conclude that no pulsed X-ray signal is detected from the Cannonball.
After taking into account the estimated background emission (except for PWN contribution) from $\S4$,
we place an upper limit on the pulse fraction at the $99.73\%$ confidence level ($3\sigma$)
of $f_p \leq 43\%$ for a sinusoidal signal in the $3-25$~keV band for the $r<20^{\prime\prime}$
aperture.

\section{Discussion}

\subsection{Is the Cannonball a Pulsar?}

\nustar\ observations of the Cannonball establish the existence of a non-thermal power-law component extending up to $\sim30$~keV. The joint fit to the
\nustar\ $-$ \chandra\ spectrum shows a single power law with a photon index of $\Gamma=1.6$ extending from 2~keV to $\sim30$~keV.  
 Furthermore, the $>10$~keV X-rays are coincident to $\sim3''$ with the central point-like 
 head of the cometary structure detected by \chandra. This is fully consistent with expectations from a PWN.  The highest energy X-rays are associated with the 
 highest energy electrons -- the ones which cool most quickly downstream from the termination shock. Given that the soft X-rays detected by \chandra\ are emanating
  from a region of $\sim2\asec$, one expects the emission above 10~keV to be from an even smaller region.  
Indeed this is consistent with the \nustar\ detection of an unresolved point source, localized on the centroid of the soft X-ray emission.
Thus the cometary soft X-ray morphology and point-like hard X-ray emission are consistent with a bulk velocity field cooled by synchrotron emission -- a PWN. 

Further support for a PWN origin for the Cannonball comes from the radio observations (ZMG13), where a cometary morphology is also detected, along 
with an extended plume and tail region consistent with a ram pressure-confined outflow. ZMG13 also point out the similarity of the Òhead-tongueÓ structure with the 
Mouse PWN. Moreover, the hardening of the radio spectrum downstream from the head-tongue structure is indicative of a cooling outflow downstream from a termination shock.  
 
The non-thermal energy spectrum for the head region of the Cannonball 
exhibits a steepening of $\Delta\alpha=\alpha_{x}-\alpha_{r}=0.9\pm0.5$, with $\alpha_{x}=\Gamma-1$.  This is comparable within uncertainties to the classic 
case of a 0.5 break in the radio-X-ray spectrum, which is the spectral steepening associated with the continuous injection of electrons into 
a homogeneous source cooled by synchrotron losses.  Further support for the PWN identification comes from the transverse velocity measured by ZMG13. 
Previous work \citep{Maeda2002,  Sakano2004, Park2005} all propose a Type II SN event as a likely origin for Sgr~A~East, and the 
$\sim500~{\rm km}$ ${\rm s}^{-1}$ transverse velocity is consistent within expectations for Type II SNe, where high pulsar ejection velocities are expected. 

Neither radio nor \nustar\ searches turn up any pulsation. It is very rare to detect radio pulsars near the Galactic Center due to 
interstellar dispersion \citep{Deneva2009}. Additionally, the Cannonball sits in a region of high X-ray background making detection of X-ray pulsations difficult. 
Our pulsed fraction upper limit of $43\%$ indicates that the Cannonball has a spin period shorter than 4~ms or its pulsar emission is buried 
under the bright PWN emission. Our result extends the range of non-detection from $147$~ms (P05) to $4$~ms.  In several young PWNe (e.g. G21.5-0.9, \citet{Nynka2013}), 
pulsar emission remains undetected in the X-ray band due to significantly brighter PWN emission. Therefore, the non-detection of pulsations from the Cannonball 
is not surprising.

Some PWN, such as G21.5-0.9, exhibit strong $\gamma$-ray emission detectable with HESS \citep{G21Hess2007}.  
Inverse-Compton emission for the Cannonball was predicted using a model by \citet{zcf2008}, incorporating an appropriate IR flux \citep{ha2007}. Based on the 
diffuse $\gamma$~ray emission in the Galactic Center \citep{viana2011}, the Cannonball would either have marginal or no detection by HESS.

\subsection{Magnetic Field of the Cannonball PWN}
 
 An estimate of the PWN magnetic field can be obtained from the \nustar\ and \chandra\ data.  
 We assume a constant (mean) magnetic field and a bulk flow velocity field downstream of the termination shock of the form $v(r) = c/3 (r_s/r)^2$ where $r_s$ is the termination shock radius. 
 These are approximations to the exact solution of \citet{Kennel1984}. Integrating this equation we obtain $r_l(E)=(c r_s^2 \tau)^{1/3}$. 
 We assume $r_l \gg r_s$,  (which the results below confirm). 
 X-ray emission ceases at downstream distance $r_l(E)$, and $\tau(E)$ is the timescale for synchrotron cooling of electrons emitting X-rays of characteristic energy $E$. 
 Normally the termination shock radius is estimated by balancing the pressure of the relativistic wind with the ram pressure of the ISM \citep{Gaensler2006}. 
 However,since the Cannonball moves through the hot plume region of Sgr~A~East, we add an additional term for the thermal pressure so that $\dot{E}/4\pi r_s^2 c \omega=\rho v^2 + P_{th}$  
 where $\dot{E}$ is the pulsar spindown power, $\rho$ the density of the ISM, $v$ the pulsar space velocity and $\omega$ a fill factor for the pulsar wind. 
P05 assume that the Cannonball outflow is energizing the plasma observed in the Plume region. 
Thus the Plume's X-ray emitting plasma provides a direct measure of the ISM density, $n$, through which the Cannonball is moving. 
From P05 one derives this ISM density for the Plume of $n\sim9~{\rm cm}^{-3} f^{-1/2}$, where $f$ is the plasma volume filling factor.
The thermal pressure of the Plume was given in P05, and we convert this to an effective thermal velocity, $P_{th} = \rho v_{th}^2 (v_{th}=465$~km~s$^{-1}$). 
The $\dot{E}$ for the pulsar cannot be obtained directly, since no pulsations were detected ($\S5$). 
However, \citet{Gotthelf2003} provides an empirical formula relating  $\dot{E}$ to the measured photon index, $\Gamma=2.36-0.021 \dot{E}_{40}^{-1/2}$.  
Using the \nustar\ derived photon index yields $\dot{E}=7\times10^{36}$~erg~s$^{-1}$ (Table \ref{tab:proptable}).  

With the above estimates the termination shock radius is $r_s = 0.006{\rm pc } (1+(v/465)^2)^{-1/2} f^{1/4} \omega^{-1/2}$, where $v$ is the pulsar space velocity in km~${\rm s}^{-1}$.  
Using the cooling time scale, a function of magnetic field strength $B$ and characteristic energy $E$ \citep{Ginzburg1965}, one can solve the cooling length scale equation to obtain  
$B=884\mu{\rm G} (1+(v/465)^2)^{-2/3} f^{1/3} \omega ^{-2/3}$. 
The cooling length scale used to derive the magnetic field was estimated from the $3-8$~keV \chandra\ image of P05, assuming 
the maximum extent of the image was associated with the (slowest cooling) 3~keV X-rays ($r_l=0.03$~pc).  
The 500~km~s$^{-1}$ transverse velocity obtained by ZMG13 is a lower limit on the pulsar space velocity. 
Approximately $90\%$ of all pulsars have space velocities less than 900~km~s$^{-1}$ \citep{Arzoumanian2002}, so using these velocities as an approximate range, 
one obtains $r_s \sim(0.003-0.005){\rm pc } f^{1/4} \omega ^{-1/2}$, consistent with estimates of the termination shock radius for other PWN. 
The magnetic field is $B \sim (313-530)~\mu{\rm G}~f^{1/3} \omega ^{-2/3}$.  
The magnetic field estimate is in excellent agreement with the lower limit of $B > 300~\mu$G obtained by ZMG13 for the Cannonball head assuming energy equipartition. 
 
\begin{deluxetable}{lcccccc}
\tablecaption{Cannonball Properties}
\tablewidth{0pt}
\tablecolumns{2}
\tablehead { \colhead{Parameters}  & \colhead{Value} }
\startdata
$L_{\rm x} (2-10$ keV) [erg s$^{-1}$]    & $6^{+2}_{-3} \times10^{33}$  \\
$L_{\rm x} (10-30$ keV) [erg s$^{-1}$]   & $7\pm1 \times10^{33}$ \\
$\dot{E}$ [erg s$^{-1}$]             	 & $\sim 7\times10^{36}$ \\
$r_s$ [pc]                               & $\sim 0.003-0.005$\\
B(PWN) [$\mu$G]                          & $\sim 313-530$ \\
B(NS) [G]                                & $\sim 5 \times 10^{12}$ \\                                                                 
\enddata                                                                                                                               
\tablecomments { Best-fit parameters of the Cannonball, joint fit with \nustar\ and \chandra\ data.  The errors are $90\%$ confidence level. 
Flux values are obtained from the power law component of the \nustar\ data. Line-of-sight distance to the Cannonball
is assumed to be 8kpc.}                                                     
\label{tab:proptable}                                                                                                                  
\end{deluxetable}     

\subsection{Magnetic Field of the Putative Pulsar}

Assuming a magnetic braking index of 3 and an initial spin period $P_{\rm o} \ll P$, where $P$ is the current spin period, an expression 
for the NS surface magnetic field strength can be obtained by (very approximately) assuming that the pulsar characteristic age is equal to the observed age $\tau$:  
$B = 3.2\times10^{19}$ G $(\pi ^{2} I/\tau ^{2} \dot{E})^{1/2}$ \citep{mt1977}. Using the $\dot{E}$ derived above, a pulsar age of 
$9000 \pm 100$ yr, and assuming 
$I=10^{45}~{\rm g}~{\rm cm}^{2}$ for a NS of $1.4 M_{\odot}$ and $R=10$~km, we obtain $B\sim5\times10^{12}$~G. 

\section{Summary}

The Cannonball has been detected by \nustar\ at energies up to $\sim30$~keV, revealing the presence of non-thermal emission.  
This observation, combined with the recent discovery of proper motion of the Cannonball with speed $\sim500$~km~s$^{-1}$ 
and direction pointing towards the center of the Sgr~A~East SNR, further solidifies the case that the Cannonball is the NS associated with Sgr~A~East. 
A timing search for pulsation was unsuccessful. An estimate of the PWN magnetic field from the X-ray data is consistent with a lower limit obtained from radio 
data by equipartition arguments.

\acknowledgements
This work was supported under NASA contract No. NNG08FD60C, and made use of data from the \nustar\ mission, a project led by the California Institute of 
Technology, managed by the Jet Propulsion Laboratory, and funded by the National Aeronautics and Space Administration. We thank the \nustar\ Operations, 
Software and Calibration teams for support with the execution and analysis of these observations. This research has made use of the \nustar\ Data Analysis 
Software (NuSTARDAS) jointly developed by the ASI Science Data Center (ASDC, Italy) and the California Institute of Technology (USA). The authors wish to 
thank Jules Halpern for useful discussions.


\end{document}